# Finding Extrasolar Planets in the Stellar Graveyard with the Hubble Space Telescope

John H. Debes[1], Steinn Sigurdsson[1], and Bruce Woodgate[2]
[1]Pennsylvania State University, University Park, PA 16803
[2]NASA Goddard Space Flight Center, Greenbelt, MD, 20771

## Abstract

*In order to directly image an extrasolar planet, the large contrast between a star and a companion planet must be overcome. White Dwarfs (WDs) are the remnants of stars > 1 $M_\odot$ and are orders of magnitude dimmer, making searches for planets and brown dwarfs (BDs) possible. We present the preliminary results of a survey of 7 hydrogen white dwarfs with photospheric metal lines (DAZs) for substellar and planetary objects, using NICMOS on the Hubble Space Telescope (HST). The extremely high contrast available with the NICMOS coronagraph allows 7-12 $M_{Jup}$ planets to be detected to within 10-30 AU of the primary WD. This represents the first concerted search for planetary objects around stars that were more massive than the sun. Our results show that there are candidates around three of the seven stars, which are awaiting confirmation through common proper motion studies, some of which can only be confirmed through another observation with HST. Continuing studies of WDs will constrain planet formation scenarios for more massive stars and provide a sample of planets to study with future space missions such as JWST and SIM.*

## Introduction

- **Direct imaging of planets around sun-like stars is not possible with current telescopes**
- **Large planets (several times the mass of Jupiter) that are less than 3 Gyr old can in principle be observed orbiting white dwarfs**
    - WDs are dimmer, which allows deeper imaging searches closer to the star
    - WDs have lost mass so companion orbits are wider than their primordial values by a factor of 2-8
    - Planets survive their star's post main sequence evolution (Duncan & Lissauer 1998)
    - **Which WDs will give the most science?**
    - We chose a sample of hydrogen WDs with photospheric metals (DAZs). The accretion of these metals may be caused by pollution due to comets or asteroids scattered inwards by a relic planetary system that survived post main sequence evolution and became dynamically unstable (Debes & Sigurdsson 2002)
    - This is not the only possible explanation, so a search will also resolve the origin of DAZs (see Dupuis et al. 1993; Zuckerman et al. 2003)
    - This search will also recover cool WD and brown dwarf (BD) companions
- **Seven targets were chosen to be observed with the Hubble Space Telescope**

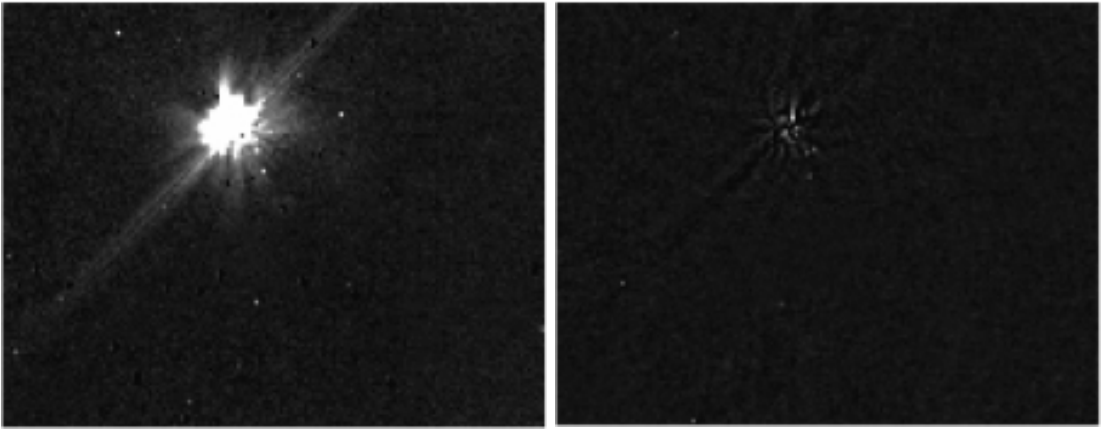

Fig. 1 (*left*) An image of G29-38 at one spcacecraft roll angle in the F110W filter ($\lambda=1.1\mu$m). (*right*) An image in the same filter but after PSF subtraction and roll angle combination. If a companion were present it would have a distinctive positive image with two negative images separated by the roll angle between the two images. The improvement in contrast close to the star is on the order of a factor of ~20.

**Observing Strategy**

- **Deep coronagraphic observations (to J~23)**
- **Shallow observations for 3 objects to look for companions blocked by coronagraph**
- **For coronagraphic images, two observations separated by a spacecraft roll (Schneider & Silverstone 2002)**
- **Used F110W (~J) and F160W (~H) filters for three coronagraphic targets, F110W for the remaining targets**
- **To confirm companionship, common proper motion will need to be measured**
    - A second epoch of observations will need to be taken with HST to confirm any nearby candidates
    - Most of our targets have proper motions on the order of .5"/year so background objects will easily be distinguished in a year or less.

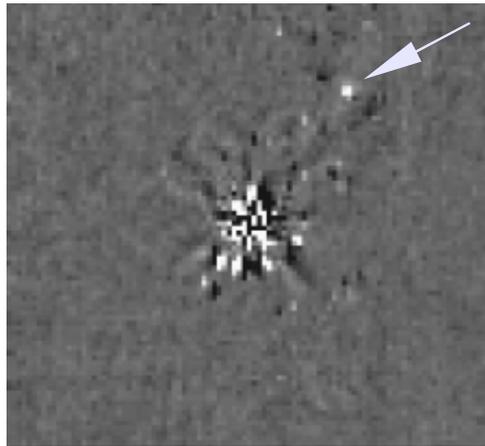

Fig 2. A candidate companion with a measured MF110W=22.3. It is separated by ~2.5". If it is associated with the host star, it would have a mass of ~15 $M_{Jup}$.

## Data Reduction and Analysis

- **We took the standard calibrated data from HST**
- **Co-added, pedestal subtracted, and registered the subexposures per spacecraft roll (Fraquelli et al. 2004)**
- **Registered and subtracted the different roll angles**
    - Two subtraction images were created, so that any companion would have a positive and negative conjugate in each image offset by the roll angle. One image was rotated and combined with the other to gain signal-to-noise.
    - Figure 1 shows the benefits of this technique, we estimate an improvement in contrast by a factor of at least 20 over the coronagraph alone, limited by readnoise
- **To check sensitivity we added artificial "companions" following the method of Schneider & Silverstone (2002)**
    - Artificial companions were added in both roll angles and considered recovered if conjugate images could be detected
    - We used the IRAF task DAOFIND to detect objects with a $5\sigma$ threshold
    - Shallower images and unsubtracted coronagraphic images were used to estimate sensitivity for separations < 0.7". This approach was more subjective as detection was determined by eye.
- **Photometry of any true companions was taken**
- **A system age was estimated from published models of WDs (Bergeron et al. 1998)**
- **Using flux models of planets, inferred mass as a function of $M_{F110W}$ and system age (Burrows et al. 2003)**
- **If candidate found at particular distance, can infer primordial separation by $a_i = (M_f/M_i) a_f$**

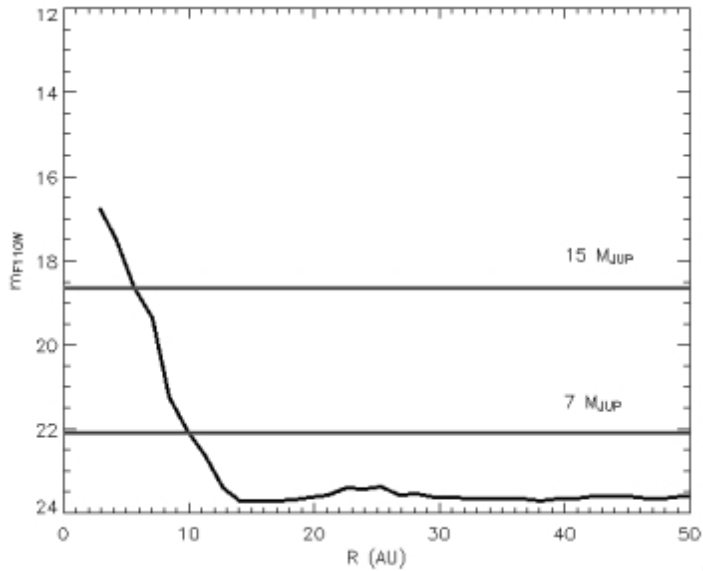

Fig4. An azimuthally averaged sensitivity profile for G29-38, derived by implanting artificial companions of varying intensity as a function of projected separation assuming a distance of 14 pc. 1"=14 AU. G29-38 has a $m_{F110W}=13.12$ and we assume a total age of 1 Gyr for the WD.

## Main Results

- **No objects > 7 $M_{Jup}$ found around G29-38 to a projected separation of 11 AU. Assuming $M_i$=3 $M_\odot$ and $M_f$=0.7$M_\odot$, we probe to a primordial semimajor axis of 2.6 AU**
- **Two candidates with projected separations < 3", one candidate at 8"**
  - The furthest candidate may be a background object but the host star is at high galactic latitude. Ground based follow up will confirm its possible association
  - The closer companions will require HST follow up
- **Candidates range in possible mass from 6$M_{Jup}$ to 25 $M_{Jup}$**
- **We estimate that < 27% of stars 2-3 $M_\odot$ have >12 $M_{Jup}$ companions primordially between 10 and 30 AU (1$\sigma$ confidence)**
  - The upper limit is based on our four youngest white dwarfs where we have the highest sensitivity. We calculated the upper limit from zero detections in four trials and assuming a Poissonian distribution
  - This estimate will change and be refined as we confirm companions and increase the sample size

## Conclusions and Future Work

- *The first extrasolar planet may have been directly imaged if common proper motion for our candidates is confirmed*
- **A larger sample of WDs should more tightly constrain the frequency of massive planets and BDs around intermediate mass stars**
- **HST is uniquely capable of conducting high contrast imaging of dim objects such as white dwarfs**
- **Future studies should include other WDs with metals to increase sample size**
- **Sifting through the stellar graveyard can allow forensic planetology to be conducted: inferring primordial conditions from present day systems.**
- **This technique could provide a sample of planetary systems that are more easily studied with spectroscopy, especially with JWST**

## References

Burrows et al. 2003 ApJ, 596, 587
Bergeron et al. 1998 ApJ, 497, 294
Debes & Sigurdsson 2002 ApJ, 572, 556
Duncan & Lissauer 1998 Icarus, 134, 303
Dupuis et al. 1993, ApJS, 84, 73
Fraquelli et al. 2004, PASP, 116, 55
Livio, Pringle & Saffer, MNRAS 257, 15
Schneider & Silverstone 2003, Proc. SPIE, 4860, 1
Zuckerman et al. 2003, ApJ, 596, 477

Support for program 9834 was provided by NASA through a grant from the STScI, which is operated by AURA, under NASA contract NAS5-26555. J.D. is also supported by a NASA GSRP fellowship through Goddard Space Flight Center, NGT5-119